\documentclass[journal]{IEEEtran} 
\usepackage{epsfig}
\usepackage{subfigure}
\usepackage{amsmath}
\usepackage{amssymb}

\begin{document}

\title{TCP throughput guarantee in the DiffServ Assured Forwarding service: what about the results?}

\author{Emmanuel Lochin$^1$ and Pascal Anelli$^2$\\
~\\
$^1$Universit\'{e} de Toulouse, ISAE, LAAS-CNRS, France\\
$^2$IREMIA - Universit\'{e} de la R\'{e}union, France \\
~\\
\small\texttt{emmanuel.lochin@isae.fr, pascal.anelli@univ-reunion.fr}
\normalsize
}
        
\maketitle
\thispagestyle{empty}

\begin{abstract}
Since the proposition of Quality of Service architectures by the IETF, the interaction between TCP and the QoS services has been intensively studied.
This paper proposes to look forward to the results obtained in terms of TCP throughput guarantee in the DiffServ Assured Forwarding (DiffServ/AF) service 
and to present an overview of the different proposals to solve the problem.
It has been demonstrated that the standardized IETF DiffServ conditioners such as the token bucket color marker and the time sliding window color maker were not good TCP traffic descriptors. Starting with this point, several propositions have been made and most of them presents new marking schemes in order to replace or improve the traditional token bucket color marker.
The main problem is that TCP congestion control is not designed to work with the AF service. Indeed, both mechanisms are antagonists.
TCP has the property to share in a fair manner the 
bottleneck bandwidth between flows while DiffServ network provides a level of service controllable and predictable.
In this paper, we build a classification of all the propositions made during these last years and compare them. As a result, we will see 
that these conditioning schemes can be separated in three sets of action level and that the conditioning at the network edge 
level is the most accepted one. 
We conclude that the problem is still unsolved and that TCP, conditioned or not conditioned, remains inappropriate to the DiffServ/AF service.
\end{abstract}


\section{Introduction}

The Differentiated Services architecture \cite{rfc2475} proposes a scalable mean to deliver IP Quality of Service (QoS) based on handling of traffic aggregates. This architecture adheres to the basic Internet philosophy namely that complexity should be relegated to the network edges while simple functionality should be located in the core network. This architecture advocates packet tagging at the edge and lightweight forwarding in the core. Core devices perform only differentiated aggregate treatment based on the marking set by the edge devices. Edge devices in this architecture are responsible to ensure that user traffic conforms to traffic profiles.  

The Assured Forwarding (AF) Per Hop Behavior (PHB) is one of the DiffServ forwarding mechanism \cite{rfc2697}. The service called Assured Service (AS) built on top of the AF PHB is designed for elastic traffics and is intended to assure a minimum level of throughput. The minimum assured throughput is given according to a negotiated profile with the client. The throughput increases as long as there are available resources and decreases when congestion occur. Such traffic is generated by adaptive applications. 

In the assured service, the throughput of these flows breaks up into two parts. First, a fixed part that corresponds to a minimum assured throughput. In the event of network congestion, the packets of this part are preserved from loss (colored green or marked in-profile). Second, an elastic part which corresponds to an opportunist flow of packets\footnote{an opportunist traffic is a traffic which occupies more bandwidth than its target rate in a congested network} (colored red or marked out-profile). No guarantee is brought to these packets. They are conveyed by the network on the principle of best-effort service (BE). In case of congestion, these packets are first dropped. This opportunistic part of the flow must vary according to the level of resources used, hence its elastic character. In any case, the throughput offered by this service must be better than the BE service.  In this architecture, the ultimate goal is to obtain an assured throughput in the absence of per-flow treatment in the network.  

The drop precedence sets in the core routers provides a good indication of the congestion level. If the network is far from being congested, the in-profile packets will rarely be dropped and their dropping probability will be neglectable. If the network is going to be congested, almost all of the out-profile packets will be dropped.  The dropping mechanism used on the core routers is generally the well-known RIO queue \cite{clark98}. RIO is the basic active queue management mechanism suitable for the setup of the AF PHB.  In order to decide whether to discard out-profile packets, respectively in-profile packets, RIO uses the average size of the total queue formed by in-profile and out-profile, respectively in-profile packets only.

Concerning the edge of the network, edge routers use a conditioner/marker in order to profile the traffic. There isn't hypothesis on the localization of these conditioner/marker. Indeed, they could be set on the client side rather than the edge router.  In the first DiffServ network specification, the edge routers used a token bucket marker mechanism in order to characterize the traffic by marking in-profile and out-profile the packets of a flow. This traffic profile consists of a minimum throughput, characterized by two token bucket parameters, namely the token rate $r$ and the size of the bucket $b$. Thus, the conformity control of an aggregate compared to the profile is done by a token bucket as proposed in \cite{rfc2697, rfc2698}.

It is likely that the assured service was designed for applications relying on the TCP protocol. TCP increments continuously its throughput and as a consequence, the bandwidth occupation by increasing the data transmission rate in function of the acknowledgement packets. If the network drops packets, TCP decreases its transmission rate. Obviously, TCP is not aware of the underlying QoS offered by the network. In the assured service, this TCP feature can involve poor performances. If a user is allowed to send packets exceeding profile requirements, these packets will be classified as out-profile by the edge routers. In case of network congestion, these packets can be dropped. Depending of the number of losses, this dropping can involve a high reduction of the transmission rate at the TCP level. As a consequence, the performance of a TCP flow carried out by the assured service is mainly determined by its out-of-profile packets. Even if the network has sufficient bandwidth for in-profile packets, the losses experienced by out-of-profile packets decrease the overall performance of the TCP flow. Indeed, the TCP congestion control is not aware of the assured traffic. This problem is the motivation of several years of research in order to correctly characterize the TCP flow in a DiffServ environment as proposed in these numerous studies \cite{EUQOS,giacomazzi03,AQUILA,GCAP,GEANT,TFTANT,TEQUILA}. In this paper, we proposes to detail these numerous proposals and to look at their impact in terms of TCP throughput guarantee over the assured service.  

\section{Background}

A network can be either \textbf{over provisioned} or \textbf{under provisioned}. Basically, these two cases deal with the excess bandwidth available in the network.

Let $r(i)_{AS}$ be the assured rate allocated to the flow $i$ (\emph{i.e.} in-profile packets throughput), 
$n$ the number of AS TCP flows in the aggregate at the bottleneck level and $C$ the link capacity. 
Precisely, this capacity corresponds to a bottleneck link in the network. If a number of $i$ flows 
cross this link, the total capacity allocated for assured service $R_{AS}$ is:

\begin{equation}
R_{AS}=\sum_{i=1}^{n}r(i)_{AS}
\label{eq:rias}
\end{equation}

Let $C_{AS}$ be the resource allocated to the assured service. If we have:

\begin{equation}
R_{AS} \leqslant C_{AS}
\label{eq:over}
\end{equation}

It means an \textbf{over provisioned} network. In this case, there is excess bandwidth in the network. 
If we are in the special case where $R_{AS} = C_{AS}$, this network is called \textbf{exactly provisioned}. 
It means there is enough bandwidth only for the in-profile traffic. 
In \cite{park04proportionnal}, the authors explain some good properties in terms of achieving a 
differentiation level with such a network. 

When: 

\begin{equation}
R_{AS} > C_{AS}
\label{eq:under}
\end{equation}

We are in the context of an \textbf{under provisioned} network:
there isn't excess bandwidth. This configuration is the worst case for the AS. It means there is available bandwidth for
the in-profile traffic only.
This service must provide an assurance until the over-subscription case is reached. Afterwards, 
the service is downgraded since no enough resources are available. As no assurance is provided, this configuration is equivalent to best effort.

In a well-dimensioned network, the inequity (\ref{eq:under}) should be avoided.
When there are losses in the network, it corresponds to the losses of out-profile packets, 
and not in-profile packets. It means that a light network congestion appears in the network and 
some out-profile packets must be dropped. 

The throughput obtained by a flow depends on the packets dropping policy of the network and
how the transport protocol reacts to these losses.
TCP reacts to a loss by halving its congestion window and increases this one linearly each time 
a packet is delivered according to the AIMD principle: \emph{additive increase} and \emph{multiplicative decrease } 
\cite{jacobson88congestion, floyd99promoting}.
 
A thorough study of the TCP and UDP behavior in the AF service was undertaken in \cite{seddigh99bandwidth}.  The latter showed 
that when the service has excess bandwidth (compared to the QoS requested), 
a flow guarantee can be given independently of these five following parameters: the Round Trip Time (RTT), the number of flows, the target rate, the size of the packets, the number of non-reactive flows (such as UDP flows).  
The distribution of the excess bandwidth between each TCP flow depends on these five parameters. 
Similar conclusions were presented in \cite{goyal99performance, rezende99assured}. 
Lastly, Seddigh et Al. \cite{seddigh99bandwidth} defines three criteria concerning equity between TCP and UDP 
according to the network state.
They show that in an over-provisioned network, all TCP and UDP flows can obtain: 1)~their target rate; 2)~a fair share of the 
excess bandwidth proportional to their target rate; 3)~in an under-provisioned network, all TCP and UDP flows observe a decrease of their throughput. 
This decrease is proportional to their assured throughput. 
Another well-known problem is that a large RTT difference between flows influences the desired assured throughput. 
In the case of identical RTT, each TCP flow in a network shares in a fair manner the available bandwidth. On the other hand, 
the TCP fair share does not exist if each flow has a different RTT.

In \cite{sahu00achievable}, Sahu et Al. demonstrates that:
\begin{itemize}
\item the obtained throughput is not proportional to the marked throughput;
\item it is not always possible to reach the target rate;
\item a flow with a high target rate will never reach its target rate if a flow with a low target rate
outperforms its profile;
\item in the case describes below, the token bucket marker parameters have no effect on the assured throughput.
\end{itemize}

Indeed, in the case of an over-provisioned network, when the loss probability of an in-profile packet can be considered has null : $p(i)_{IN } = 0$ 
and that the loss of an out-profile is not : $p(i)_{OUT} > 0$, if the target rate of a flow verifies
the following equation :   
\begin{equation}
r(i)_{AS} < \frac{1}{RTT} \sqrt{\frac{3}{2~p(i)_{OUT}}}
\label{eq:influence}
\end{equation}
then the token bucket marker has no effect on the reached throughput \cite{sahu00achievable}.

This important result gives a strong limitation to the use of the token bucket marker for TCP conditioning. Indeed, equation (\ref{eq:influence}) shows that a simple token bucket marker is unable to achieve a large range of requested target rate by increasing or decreasing the out-profile marking of a TCP flow.
As a result, new marking strategies propose to control the TCP achieved throughput by dynamically choosing a target rate $r(i)_{AS}$
as a function of the loss ratio.

\section{Marking strategy big picture}

\begin{figure}
\begin{center}
\subfigure[TCP flows with different RTT]{\epsfig{figure=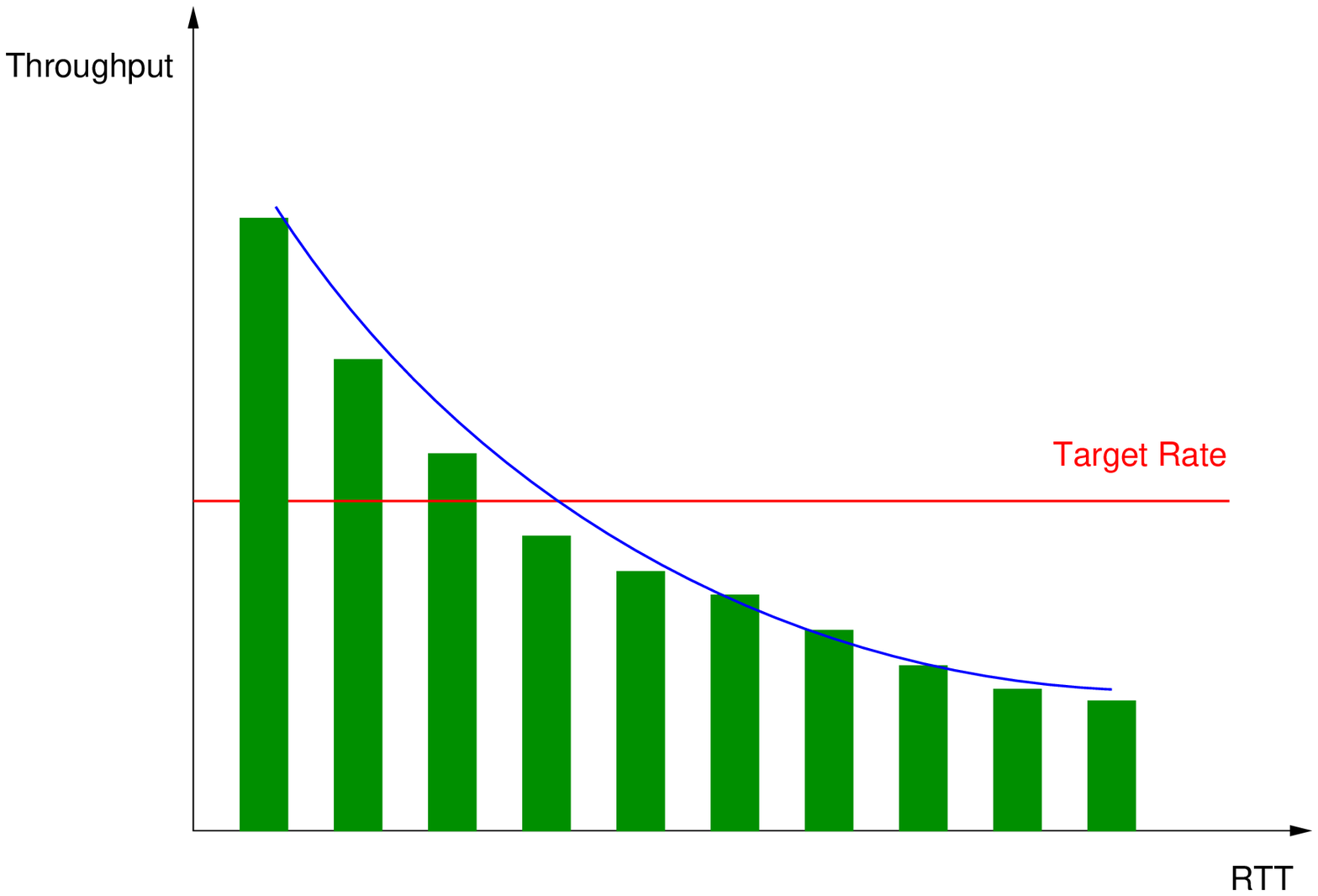, width=4cm}}
\subfigure[Excess and deficit areas]{\epsfig{figure=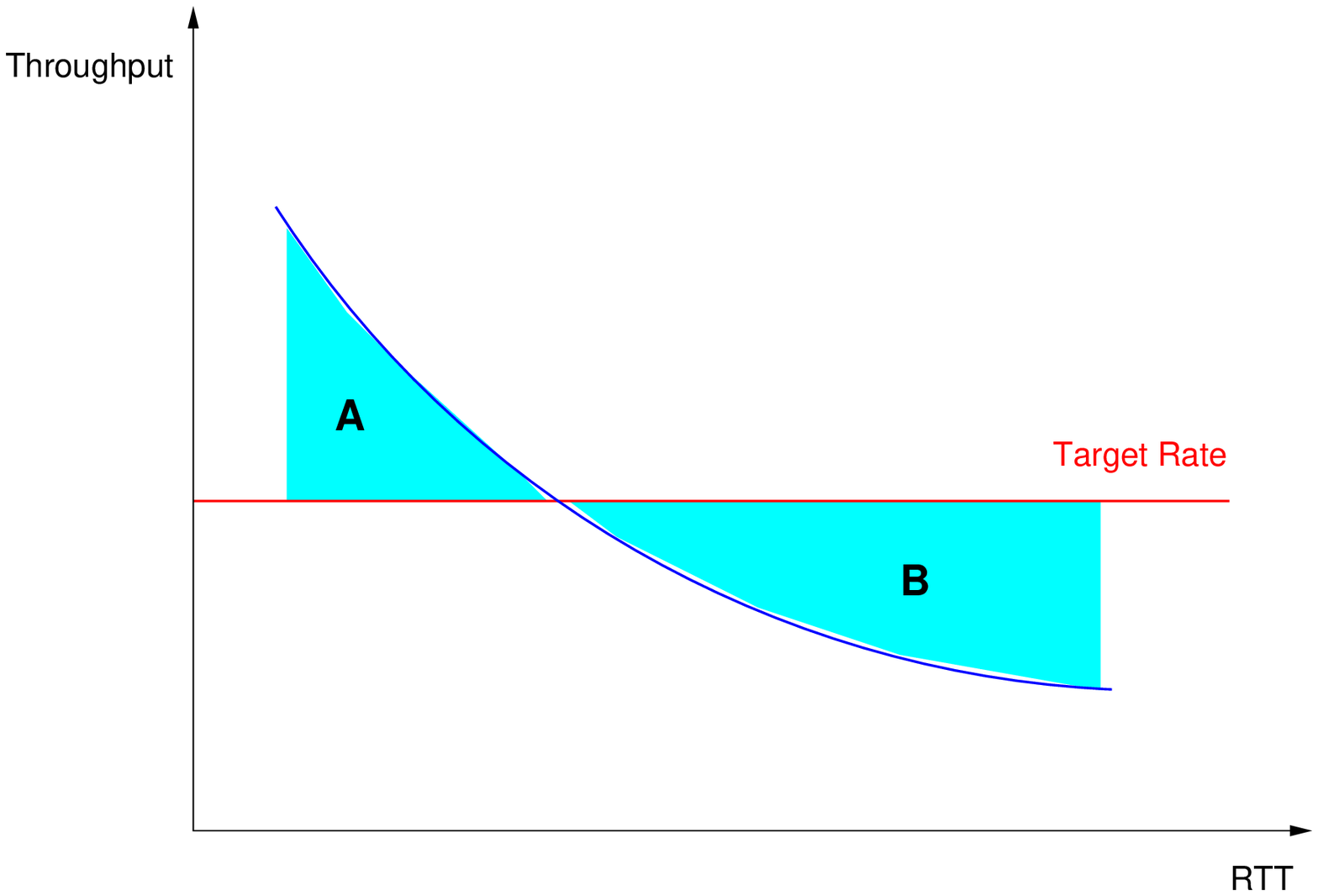, width=4cm}}
\caption{TCP throughput as a function of RTT}
\label{fig:rtt-aire}
\end{center}
\end{figure}

Basically, the principle of the marking strategy is to infer on the TCP throughput of the opportunist flow by controlling the
number of losses in their out-profile parts. Following the simple model of the TCP throughput given in \cite{mathis97macroscopic}:

\begin{eqnarray}
\text{TCP throughput} = \frac {C * MSS}{RTT * \sqrt{p}}
\label{eq:tcp}
\end{eqnarray}

With $C$ a constant and $p$ the loss probability and $MSS$ the maximum segment size. 
In order to increase the loss probability of the opportunist flows, almost all the DiffServ conditioners presented 
in section \ref{sec:state} are based on the increase of the out-profile part of theses flows. 
As a result, the loss probability raises and the TCP throughput decreases as shown in (\ref{eq:tcp}). 
Unfortunately, changing the $p$ value from the equation (\ref{eq:tcp}) thanks to a marking strategy is complex.
Indeed, it is necessary to evaluate the loss probability of the network and to estimate an RTT
for each flow. In order to obtain these keys values, the authors in \cite{gendy02ebm} propose to compute an average loss
interval thanks to the method presented in \cite{floyd00equationbased} instead of the loss probability
and estimate the RTT with
a time stamping method. On the other hand, the authors in \cite{habib02round} proposes an RTT-RTO conditioner that is
based exclusively on throughput measurement and in addition of RTT estimation, proposes a solution
to take into account the TCP timeout in the marking strategy.
In order to illustrate the probability marking concept, figure \ref{fig:rtt-aire} presents the aim 
of this marking strategy. Figure \ref{fig:rtt-aire}~(a) symbolizes the throughput obtained by ten flows with
different RTTs and the same target rate. The smaller the RTT, the higher is the throughput. 
The flows with a small RTT occupies more bandwidth than necessary as explained
by area A in figure \ref{fig:rtt-aire}~(b).
Basically, the aim of a marking strategy is to distribute fairly excess bandwidth from area A to area B.
In order to summarize, we identify these three important points:
\begin{itemize}
\item the TCP throughput is closely related to the packet loss probability, the RTT and the target rate;
\item the loss of an out-profile packet is always prejudicial to the TCP assured throughput;
\item the loss probability, RTT and RTO estimations are complex to estimate in a passive manner (.i.e in an intermediate node and not 
at the sender side).
\end{itemize}

\section{Synthesis of the methods used to obtain a TCP throughput guarantee}

In the DiffServ architecture, we can act on three different levels to solve the TCP throughput guarantee problem in the AF service: at the hosts level, at 
the edge of the network or at the core network inside the Active Queue Management (AQM).
\begin{itemize}
\item at the TCP level: solutions suggested raise some deployment problems. First of all, it needs a modification of the TCP code.
This is a problem in regard of the numerous diversity and versions of the operating systems
and the number of hosts in the Internet. In the context of a DiffServ architecture, the marking is carried out exclusively by the source.
In this case, marking is not under the responsibility of the Internet Service Provider (ISP).
The checking of the marking by the ISP is not either without raising difficulties of realization. 
Lastly, this solution is not possible when marking is carried out on the aggregate.  

In \cite{feng98adaptive}, an evolution of the TCP congestion control proposes to integrate the marking according to a profile. 
The solution consists in splitting the congestion into two parts: one for each part of the assured service.
The size of each part of the congestion window changes depending on the 
network state and the observed throughput. Thus, the marking probability is computed from the assured part of the congestion window;
\item at the conditioning level:
the objective is to copy a marking which is in conformity with the TCP dynamics. Marking is a functionality which should remain under the responsibility 
of the ISP. Conditioning is an element which is put on cut on the road.  It can evolve and move independently of the other components of the 
DiffServ architecture;
\item at the AQM level, 
new scheduling techniques such JoBS \cite{chliab02a} makes it possible to impose flows guarantees in the assured service.  
These techniques are derived from the proportionality mechanisms introduced by \cite{dovrolis00proportional}.  
Another solution would be in the inter-dealing which the AQM with a TCP source can have. The \textit{Explicit Congestion Notification} TCP flag
is often used as a complement to control the throughput of a flow in order to limit the packets marked out-profile in the network.
\end{itemize}

In the following, these three levels of action are compared to the guarantee provided to the flow, the facility of deployment, the scalability.  
The most tackled solution is on the conditioning level. The prolific literature is an illustration.
Nevertheless, we will see that the solution is not obvious and that this level is not inevitably the good one.  

\section{State of the art of the TCP conditioning}
\label{sec:state}
This section gives an extended overview of the concepts used inside the DiffServ conditioners and illustrates the mechanisms
chosen to achieve the desired target rate. It will not be interesting to describe all the existing solutions as some 
of them deal with similar approaches. So, we have selected a set of DiffServ conditioners in order to highlight the concepts
used to solve the TCP throughput guaranteed problem and selected some AQM which tries to enforce the service differentiation. 

Concerning the TCP marking proposals, they are divided into two families: those which treat the TCP marking with an aggregate profile, 
and those which treat TCP marking compared to an individual profile. The former aims the equity in addition to the flow guarantee 
sought by the latter.   
We will show that most of these solutions are based on the time sliding window algorithm and/or on the two or three token bucket 
colors marker. 
Moreover, we will see that the majority of these approaches are based on a weighted probabilistic marking of the excess traffic.

\subsection{Proportional Differentiated Services}

This proposal, presented in \cite{dovrolis00proportional}, doesn't deal with a marking strategy as the service differentiation is
made at the AQM level. However, since this scheme inspired many proposals in the field of the TCP throughput guaranteed in the DiffServ/AF service, 
we present in this section the concept introduced in \cite{dovrolis00proportional}. In this proposal, each packet arriving in the network is marked either in-profile or out-profile according to a token bucket marker.  
It is on the AQM level, within the router, that the treatment is carried out. Assume two flows with two target rates  $r_1$ and $r_2$ having an 
RTT and an identical packets size. On the basis of the TCP equation (\ref{eq:tcp}), we have:

\begin{equation}
r_i \leqslant \frac{1.5 \sqrt{\frac{1}{3}} * k_i}{RTT * \sqrt{p_i}}
\label{eq:floyd}
\end{equation}
With :
\begin{itemize}
\item $r_i$: target rate of flow $i$;
\item $k_i$: packets size of the $i$ flow;
\item $p_i$: loss probability of the $i$ flow.
\end{itemize}
following the equation (\ref{eq:floyd}), we obtain:
\begin{equation}
\frac{r_1}{r_2} =  \sqrt{\frac{p_2}{p_1}}
\end{equation}
If we compare the number of dropped packets time unit: $d1$ and $d2$, corresponding to the losses throughput, we obtain:
\begin{equation}
\frac{d_1}{d_2}=\frac{r_1*p_1}{r_2*p_2}=\sqrt{\frac{p_1}{p_2}}=\frac{r_2}{r_1}
\end{equation}
It means that the number of dropped packets per time unit must be inversely proportional
to the target rate of a flow. This concept of proportionality is the basis of many studies in the TCP 
throughput guarantee such \cite{chliab02a}.
It paves the first step on dropping based on the desired target rate and will be derived and enhanced.

\subsection{Qualitative microflows marking \cite{mellia03}}

In an other hand, Marco Mellia in \cite{mellia03} study the feasibility of improving the performance of TCP flows in a network
with RIO routers by marking packets according to per-flow TCP states at network edges.
The key observation is that TCP performance decreases significantly either in the presence of bursty, non adaptive
cross-traffic or when it operates in the small window regime, i.e., when the congestion
window is small. This is because bursty losses or losses during the small window regime may cause
retransmission timeouts (RTOs), which will ultimately result in TCP entering the slowstart
phase. The objective of the TCP-aware marking algorithm is then to selectively mark packets in
order to reduce the possibility of TCP entering these undesirable states.
Marco Mellia exploits the fact that IN packets are delivered with a very high probability. Thus selectively
marking packets as IN allows TCP to exit as fast as possible from the undesirable states.
In order to take into account these states, Marco Mellia in \cite{mellia03} proposes:
\begin{itemize}
\item to mark the first several packets of the flow. This will protect the first packets against loss, and
it will allow TCP to safely exit the initial small window regime;
\item to mark several packets after an RTO occurs. The purpose of this is to make sure that the retransmitted
packet is delivered with high probability, and that TCP sender exits the small window regime
which follows the Slow Start phase entered after the RTO event;
\item to mark several packets after receiving three  duplicate acknowledgements. The present idea is to protect the 
retransmitted packet in order to allow TCP to come out the Fast Recovery phase without losing other packets.
\end{itemize}

This marking scheme is qualitative as it can improve the throughput of long lived TCP flows up to 20\%,
and reduce the completion time of short lived TCP flows by half according to the author.
The main disadvantage of this approach is that it needs to know the TCP window size and the slow-start threshold 
(\emph{ssthresh}). So, it needs to operate to a modification of the TCP stack in order to use this algorithm.

This algorithm improves a target rate but does not give any guarantee about the target rate requested.
 
\subsection{Marking schemes based on the \emph{Time Sliding Window} algorithm}

Several algorithms of this type were proposed to work with the AS service. The Two Rate Three Color Marker (TRTCM) \cite{rfc2698} based 
on a token bucket estimator algorithm and the Time Sliding Window Three Color Marker (TSW3CM) \cite{rfc2859} based on 
an average throughput estimator: the Time Sliding Window (TSW) \cite{clark98}.  In these markers, two rates are defined: an assured rate 
called Committed Information Rate (CIR) and a maximum rate: the Peak Information Rate (PIR) used in case of excess bandwidth.  

The main difference between these two markers is the way they mark the packets. Even if they take each one in argument the assured rate:  
$r(i)_{AS}$, at the opposite of the TRTCM, the TSW3CM applies a probabilistic packets marking. Indeed, the TRTCM marks a packet out-profile
if this one is not in the profile defined by the \emph{token bucket} parameters: $(r, b)$.  
On the other hand, the TSW3CM marks a packet out-profile with a probability as a function of the average rate estimated 
by the TSW and the PIR and CIR. The TSW3CM gives better results than a simple TRTCM as its marking scheme describes better the TCP traffic.
Starting with this point, a lot of others marking strategies proposed to improve this static probability marking scheme. 

We define these enhanced marking strategies as adaptive marking. In the following, we present in details enhanced proposals based on 
the TRTCM and TSW3CM.

\section{Towards an enhanced TCP conditioner: the Adaptive Marking}

The adaptive marking proposes to improve the TRTCM or the TSW3CM conditioners by changing dynamically
the marking rate. It means that the target rate of the marker evolving in the time as a function of the
network conditions and the throughput obtained by a conditioned flow.
We give below an overview of three majors adaptive algorithms.

\subsection{Adaptive marking with dynamic target rate}

In \cite{yeom01adaptive}, Yeom and Reddy present a marking scheme for a TCP flow inside an aggregate. 
This scheme is based on a mathematical TCP model defined in \cite{yeom01modeling}. This model is given 
in equation (\ref{eq:tcp-yeom}). 

\begin{equation}
b_i = \frac{3}{4}m_i+\frac{3 k_i}{4 RTT}\sqrt{\frac{2}{p_i}}
\label{eq:tcp-yeom}
\end{equation}

assume that:

\begin{equation}
b_i = \frac{3}{4}m_i+\epsilon _i
\label{eq:tcp2-yeom}
\end{equation}

With $b_i$: throughput of the $i$ flow; $k_i$: its packets size ; $p_i$: its loss probability
and $m_i$: its initial target rate value which corresponds to the $r(i)_{AS}$ token bucket marker parameter.
This equation gives the throughput of the flow as a function of the token bucket marker parameters used.
Thanks to the equation (\ref{eq:tcp2-yeom}), Yeom and Reddy propose to act on the marking process as a function of
the following states:
\begin{enumerate}
\item if $b_i \leqslant \frac{3}{4}m_i+\epsilon _i < r(i)_{AS}$: 
in this state, the flow observes an oversubscribed network, and some in-profile packets are lost. Thus,
the marker reduces $m_i$ so that $b_i$ is maintained to be higher than $\frac{3}{4}m_i$ to avoid wasting resources;
\item if $\frac{3}{4}m_i+\epsilon _i < b_i < r(i)_{AS}$: 
in this state, the flow does not reach its target. Since the network is not oversubscribed, $b_i$ can
be increased by increasing $m_i$. Thus, the marker increases
$m_i$ of that flow if resources are available;
\item if $r(i)_{AS} \leqslant b_i$: in this state, the flow already achieved its target.
Thus, the marker reduces $m_i$ to avoid wasting resources.
\end{enumerate}

In \cite{chait01providing}, Chait et Al. present a similar concept with a dynamic token bucket marker configuration.
The constituent components of this design include two-color token bucket edge markers 
coupled with a two-level AQM controller embedded in the core routers. The interactions 
between TCP flows and these components are managed by a proportional-integral controller (PI) which is a control loop feedback 
mechanism widely used in automatic and control systems. 
The PI controller attempts to adjust the target rate of the TCP flows as a function of the information returned by the network and
the current TCP achieved throughput.

These mechanisms, based on a dynamic target rate parameter, do not provide a fair sharing of excess
or lack of bandwidth when the network is respectively over-subscribed or
under-subscribed. Indeed, the allocation is determined
by the dynamics of the TCP congestion control mechanism. In the following sections, we present
an extension of this dynamic approach but with a fair sharing of the excess bandwidth.

\subsection{Adaptive marking based on memorisation}

This conditioning based on memorisation was proposed in \cite{kumar01amemory}. The principle of marking inherits from the TSW3CM
algorithm except that the marking probability is weighted by the use of a variable memory. This variable keeps a history of the
average throughput estimated by the TSW algorithm of the TSW3CM marker. It is used for indirectly detecting a variation of the 
TCP window size or an RTT variation of the conditioned flow. This method improves the fair sharing of the excess bandwidth between 
the flows whatever their RTT or their target rates.

\subsection{Adaptive marking based on a marking probability}

In the previous section, we saw that Yeom et Al. used a mathematical model of a TCP flow in a DiffServ network in order to obtain 
a theoretical value of the TCP throughput. They use this model to act on the marking rate of the token bucket marker. 
The advantage of this approach is that it can operate with a conditioning method based on the microflow or the aggregate level.  
However, this solution is not efficient in all network conditions as the model doesn't take into account all 
the network and TCP parameters (such as the TCP timeout value, the $RTT$ variation, ...).
On the other hand, in \cite{gendy02ebm}, Gendy showed that it exists a duality between the marking based on the evaluation 
of the throughput and the loss probability. As a result, Gendy proposes to find the best marking rate according to 
a more accurate TCP model proposed in \cite{padhye98modeling}. The scheme has a better feedback of the network 
state than \cite{yeom01adaptive} since it takes into account and evaluates all the parameters of this complex and accurate model.
However, this scheme is strongly limited to the accuracy of the passive measurements used to feed this equation.

The equation used is given in \cite{padhye98modeling} (we denote $F()$ this equation, see appendix A for details) and takes into account: 
   $p$: the packet loss probability;
   $W_{max}$: the maximum TCP window size;
   $RTT$: the round trip time;
   $RTO$: the TCP timeout value;
   $MSS$: the maximum segment size;
and returns the throughput $X$ with:

\begin{equation}
X = F(p_{OUT}, Wmax, RTT, RTO, MSS)
\end{equation}

Basically, if we assume that we are in a well-provisioned network (i.e. equation (\ref{eq:over}) is true). The loss probability of a packet is corresponding to the loss probability of an out-profile packet:
$p_{OUT}$ since the loss probability of an in-profile packet should be near zero: $p_{IN} \simeq 0$. 
The main difficulty is to reverse this equation in order to obtain $p$ as a function of $X$:

\begin{equation}
p_{OUT} = F(X, Wmax, RTT, RTO, MSS) 
\end{equation}

by changing $X$ with the target rate of a flow ($r_i$) we obtain: 

\begin{equation}
p_{OUT} = F(r_i, Wmax, RTT, RTO, MSS)
\end{equation}

The idea is to solve this equation in order to obtain the optimal marking rate $r(i)_{AS}$ of the token bucket parameter.

Although this proposal is certainly the most achieved one, the complexity induces by the passive measurements used to feed the equation, the need of a strong accuracy and the per-flow monitoring involved in the conditionning process are the main barrier to the deployement of such mechanism. In the following section \ref{sec:dealecn}, we will see how other proposals have overcome the problem of assessing the network by using the Explicit Congestion Notification (ECN) mechanism \cite{rfc3168}.

\section{Dealing with ECN feedbacks}
\label{sec:dealecn}

Following the difficulty to feed the parameters used to characterize the TCP flows, recent approaches
propose to use ECN feedbacks in order to assess the congestion level in the network. The main idea is
to use either the number of ECN marked packets or specific information carried inside the feedback packet as 
a congestion indication level.

\subsection{Proportional Bandwidth Allocation}

In \cite{park04proportionnal}, Park and Choi analyze the steady state throughput of TCP flows in a differentiated network. They show that
current DiffServ networks are biased in favor of those flows that have a smaller target rate, which results in an unfair bandwidth
allocation. However, they demonstrate when the network is exactly-provisioned, there is no bias in favor of an aggregate that has a smaller 
target rate. So, they propose to adjust the target rate of the token bucket markers in order
to match the bottleneck capacity as a function of the network congestion level. In other words, the sum of each marking rates should be equal
to the bottleneck capacity, this result in having almost in-profile traffic in the network.
This approach is original since it deals with two new concepts. First, the amount of ECN marked packet drives the target rate value and
second, having a network exactly-provisioned should reduce TCP sources oscillation as the number of dropped packets decreases.

Unfortunately, this solution is strongly linked to the RTT of each flow. In their paper, the authors evaluate their solution with RTT equals
or in the same order of magnitude. As a consequence, this solution cannot be generalized to a multi-domain network with a large range of RTT values.

\subsection{AIMD Penalty Shaper}

Finally, as opposed to the marking strategy adopted by new conditioners, we have proposed a delay based shaper \cite{lochin05penalty}. 
This shaper applies a delay penalty to a flow if there are out-profile packets losses in the network 
and if it outperforms its target rate. The basic idea is that the penalty is a function of the out-profile packet losses. 
Instead of raising the $p$ value, from equation (\ref{eq:tcp}), 
of the most opportunist flow, the AIMD Penalty Shaper raises a delay penalty to the flow. It results in a growth of the RTT.
Mathematically, as shown in (\ref{eq:tcp}), increasing $RTT$ value is similar to increasing $p$ value in terms of TCP throughput.
In \cite{yeom99realizing}, the authors have shown that limiting out-profile packets is a good policy to achieve a target rate.
Indeed, by limiting packets dropping we avoid TCP retransmission. 
This is an efficient solution to optimize the bandwidth usage.
Thus, the goal is to reduce out-profile losses by applying a delay penalty to the flows that are
the most opportunist in the network.
Therefore, when a RIO\footnote{RED with IN and OUT} \cite{clark98} router in the core network 
is dropping out-profile packets, it marks the
ECN flag \cite{rfc3168} of the in-profile packets en-queued in the RIO queue.
In a well-dimensioned network, there is no in-profile packet loss. Then,
the edge device can be aware that there is a minimum of one flow, or set of flows, which are opportunists in the network.  This opportunist traffic
is crossing the same path. The edge device evaluates its 
sending rate thanks to a Time Sliding Window (TSW) algorithm \cite{rfc2859}. If its sending
rate is higher than its target rate, it considers that its traffic may be opportunist. 
Then, it applies a penalty to the incoming traffic
until the network returns that there are out-profile packets losses.
This penalty allows a raise of the RTT and consequently, decrease the TCP throughput.
In \cite{lochin05aimd}, the authors choose to use an AIMD penalty instead in order 
to decrease rapidly the throughput \cite{rfc2914}. 
If there is no loss anymore, the penalty decreases linearly and the TCP throughput raises.
This principle follows the TCP congestion control.
The main advantage of this solution is that the conditioning can be made on flows with similar RTTs (i.e. in the same 
order of magnitude). Moreover, this solution doesn't depend on the complex problem of RTT estimation necessary 
to the functioning of the conditioners presented before.

\section{Discussion}
\label{sec:conclusion}

Among the multiple conditioning schemes presented, it resides two main classes. First, 
the quantitative conditioning class which includes: 
equation based marking \cite{padhye98modeling, gendy02ebm, yeom01adaptive}~;
memory-based marking \cite{kumar01amemory}~;
TSW-based marking \cite{rfc2698, clark98} and penalty shaper conditioning \cite{lochin05penalty, lochin05aimd}.
Second, the qualitative conditioning class with the marking scheme inspired by \cite{mellia03}.

If these conditioning mechanisms work well theoritically or in simulation, the scalabity of most of these proposals is not proofed in particular in case of real world experiments with cross-traffic and this case can strongly
decrease the efficiency of many conditioners. Furthermore, even if the DiffServ architecture is based on per-flow 
conditioning, it is obvious that for an ISP, it will be more easier to profile a client emitted TCP aggregate than 
every single TCP flows of a TCP source.   
Indeed, the client is typically a source domain as defined in \cite{rfc2475} that communicates with another source domain within a DiffServ core network. A good conditioner should provide a service differentiation between two source domains, 
on a set of TCP flows, based on its marking profile. 

In the context of the use of ECN feedbacks, to the best of our knowledge, it exists no analytical study in order to assess the network congestion level as a function of the amount of ECN traffic. A recent study from Bob Briscoe \cite{briscoe07reecn} proposes to extend the ECN protocol in order to
carry a better and truthful prediction of the congestion of the path. However, there is still no rule allowing to compute in an accurate manner the exact congestion level of a network following the number of RED-ECN traffic marked.

Another widespread idea is to claim that the over-provisioning is the best solution allowing the DiffServ/AF service to work without any kind of improvements. However, this method has a cost and even in case of over-provisioned network, there is no 
better guarantee to reach and maintain a negotiated target rate. The question remains the same: how to size this 
provisionning at a low cost? Indeed, there is no exact method to determine efficiently the necessary excess bandwidth size.

In 2001, in an interesting and well-known unpublished technical report \cite{firoiu01advances}, Victor Fioriu and Al. wrote that: \emph{``the possibility to use TCP in order to provide differentiated QoS is under question and replacing the TCP congestion mechanism in the context of QoS networks is currently an open research area''}. Nowadays, new approaches have proposed to design specific DiffServ transport protocols (such as \cite{ernesto04design}, \cite{jourjon08qstp}) able to be aware of the negotiated QoS thanks to a QoS congestion control and cross-layer mechanisms allowing the transport protocol to be fully aware of the target rate negotiated between the application an the network provider. Following these proposals which have demonstrated their complete compliances with DiffServ QoS network architecture such as the EUQoS\footnote{http://www.euqos.eu/} architecture \cite{jourjon06implementation}, the question to use TCP as QoS transport protocol in order to provide services guarantees seems outdated.  

\appendix{A. TCP Model}
$$F(p_{OUT}, W_{max}, RTT, RTO, MSS)=$$ 
\tiny
\begin{equation}
\left\{ \begin{array}{ll}
MSS \frac{ \frac{1-p}{p} + W(p) + \frac{Q(p,W(p))}{1-p}} {RTT (\frac{b}{2} W(p) + 1) + \frac{Q(p,W(p)) F(p) To}{1-p}} & \mbox{if $W(p) < W_{max}$} \\
MSS \frac{ \frac{1-p}{p} + W_{max} + \frac{Q(p,W_{max})}{1-p}} {RTT (\frac{b}{8} W_{max} + \frac{1-p}{p W_{max}} + 2 ) + \frac{Q(p,W_{max}) F(p) To}{1-p}} & \mbox{otherwise} \\
\end{array}
\right.
\end{equation}
\normalsize
where
\tiny
\begin{eqnarray}
W(p) & = & \frac{2+b}{3b} + \sqrt{\frac{8(1-p)}{3bp} + {\frac{2+b}{3b}}^{2}} \\
Q(p,w) & = & min \left( 1,\frac{ (1-{(1-p)}^{3})(1+{(1-p)}^{3}(1-{(1-p)}^{w-3}) }{1 - {(1-p)}^{w}} \right) \\
F(p) & = & 1 + p + 2{p}^{2} + 4{p}^{3} + 8{p}^{4} + 16{p}^{5} + 32{p}^{6}
\end{eqnarray}
\normalsize
$b:$ is the average number of packets acknowledged by an ACK, usually 2.

\bibliographystyle{plain}
\bibliography{../bib/biblio}

\begin{thebibliography}{10}

\bibitem{rfc2475}
S.~Blake, D.~Black, M.~Carlson, E.~Davies, Z.~Wang, and W.~Weiss.
\newblock An architecture for differentiated services.
\newblock Request For Comments 2475, {IETF}, December 1998.

\bibitem{briscoe07reecn}
Bob Briscoe, Arnaud Jacquet, Alessandro Salvatori, Martin Koyabe, and Toby
  Moncaster.
\newblock {Re-ECN}: {A}dding accountability for causing congestion to {TCP/IP}.
\newblock Internet Draft draft-briscoe-tsvwg-re-ecn-tcp-04.txt, {IETF}, July
  2007.

\bibitem{chait01providing}
Y.~Chait, C.~Hollot, V.~Misra, D.~Towsley, and H.~Zhang.
\newblock Providing throughput differentiation for {TCP} flows using adaptive
  two color marking and multi-level aqm.
\newblock In {\em Proc. of {IEEE INFOCOM}}, New York, June 2002.

\bibitem{chliab02a}
N.~Christin, J.~Liebeherr, and T.~Abdelzaher.
\newblock A quantitative assured forwarding service.
\newblock In {\em Proc. of {IEEE INFOCOM}}, volume~2, pages 864--873, New York,
  NY, June 2002.

\bibitem{clark98}
D.~Clark and W.~Fang.
\newblock Explicit allocation of best effort packet delivery service.
\newblock {\em IEEE/ACM Transactions on Networking}, 6(4):362--373, August
  1998.

\bibitem{rezende99assured}
José~Ferreira de~Rezende.
\newblock Assured service evaluation.
\newblock In {\em Proc. of {IEEE GLOBECOM}}, pages 100--104, Rio de Janeiro,
  Brasil, December 1999.

\bibitem{dovrolis00proportional}
C.~Dovrolis and P.~Ramanathan.
\newblock Proportional differentiated services, part ii: Loss rate
  differentiation and packet dropping.
\newblock In {\em Proc. of IEEE/IFIP International Workshop on Quality of
  Service - IWQoS}, Pittsburgh, PA, June 2000.

\bibitem{gendy02ebm}
M.A. El-Gendy and K.G. Shin.
\newblock {Assured Forwarding Fairness Using Equation-Based Packet Marking and
  Packet Separation}.
\newblock {\em Computer Networks}, 41(4):435--450, 2002.

\bibitem{ernesto04design}
{Ernesto Exposito and Michel Diaz and Patrick S{\'e}nac}.
\newblock {Design Principles of a QoS-oriented Transport Protocol}.
\newblock In {\em {IFIP International Conference on Intelligence in
  Communication Systems}}, Bangkok, November 2004.

\bibitem{EUQOS}
EuQoS.
\newblock End-to-end quality of service support over heterogeneous networks.
\newblock http://www.euqos.org/.

\bibitem{rfc2859}
W.~Fang, N.~Seddigh, and AL.
\newblock {A Time Sliding Window Three Colour Marker}.
\newblock Request For Comments 2859, {IETF}, June 2000.

\bibitem{feng98adaptive}
W.~Feng, Dilip Kandlur, Debanjan Saha, and Kang~S. Shin.
\newblock {Adaptive Packet Marking for Providing Differentiated Services in the
  Internet}.
\newblock In {\em Proc. of the IEEE International Conference on Network
  Protocols - ICNP}, October 1998.

\bibitem{firoiu01advances}
V.~Firoiu, J.~Le Boudec, D.~Towsley, and Z.~Zhang.
\newblock Advances in internet quality services.

\bibitem{rfc2914}
S.~Floyd.
\newblock Congestion control principles.
\newblock Request For Comments 2914, {IETF}, September 2000.

\bibitem{floyd99promoting}
Sally Floyd and Kevin Fall.
\newblock Promoting the use of end-to-end congestion control in the {Internet}.
\newblock {\em IEEE/ACM Transactions on Networking}, 7(4):458--472, 1999.

\bibitem{floyd00equationbased}
Sally Floyd, Mark Handley, Jitendra Padhye, and Jorg Widmer.
\newblock {Equation-based Congestion Control for Unicast Applications}.
\newblock In {\em Proc. of {ACM SIGCOMM}}, pages 43--56, Stockholm, Sweden,
  August 2000.

\bibitem{goyal99performance}
M.~Goyal, A.~Durresi, R.~Jain, and C.~Liu.
\newblock Effect of number of drop precedences in assured forwarding.
\newblock In {\em Proc. of {IEEE GLOBECOM}}, pages 188--193, 1999.

\bibitem{jourjon08qstp}
Emmanuel~Lochin Guillaume~Jourjon and Patrick Senac.
\newblock Design, implementation and evaluation of a qos-aware transport
  protocol.
\newblock {\em To appear in Computer Communications}, 2008.

\bibitem{habib02round}
Ahsan Habib, Bharat Bhargava, and Sonia Fahmy.
\newblock {A Round Trip Time and Time-out Aware Traffic Conditioner for
  Differentiated Services Networks}.
\newblock In {\em Proc. of the IEEE International Conference on Communications
  - ICC}, New-York, USA, April 2002.

\bibitem{rfc2697}
J.~Heinanen and R.~Guerin.
\newblock {A Single Rate Three Color Marker}.
\newblock Request For Comments 2697, {IETF}, September 1999.

\bibitem{rfc2698}
J.~Heinanen and R.~Guerin.
\newblock A two rate three color marker.
\newblock Request For Comments 2698, {IETF}, September 1999.

\bibitem{jacobson88congestion}
Van Jacobson.
\newblock Congestion avoidance and control.
\newblock In {\em Proc. of {ACM SIGCOMM}}, pages 314--329, Stanford, CA, August
  1988.

\bibitem{jourjon06implementation}
Guillaume Jourjon, Emmanuel Lochin, Laurent Dairaine, Patrick Senac, Tim Moors,
  and Aruna Seneviratne.
\newblock {Implementation and performance analysis of a QoS-aware TFRC
  mechanism}.
\newblock In {\em Proc. of {IEEE ICON}}, Singapore, September 2006.

\bibitem{kumar01amemory}
K.~Kumar, A.~Ananda, and L.~Jacob.
\newblock {A Memory based Approach for a TCP-Friendly Traffic Conditioner in
  DiffServ Networks}.
\newblock In {\em Proc. of the IEEE International Conference on Network
  Protocols - ICNP}, Riverside, California, USA, November 2001.

\bibitem{lochin05aimd}
Emmanuel Lochin, Pascal Anelli, and Serge Fdida.
\newblock {AIMD Penalty Shaper to Enforce Assured Service for TCP Flows}.
\newblock In {\em Proc. of the International Conference on Networking - ICN},
  La Reunion, France, April 2005.

\bibitem{lochin05penalty}
Emmanuel Lochin, Pascal Anelli, and Serge Fdida.
\newblock Penalty shaper to enforce assured service for {TCP} flows.
\newblock In {\em IFIP Networking}, Waterloo, Canada, May 2005.

\bibitem{mathis97macroscopic}
M.~Mathis, J.~Semke, and J.~Mahdavi.
\newblock {The Macroscopic Behavior of the {TCP} Congestion Avoidance
  Algorithm}.
\newblock {\em Computer Communications Review}, 27(3), 1997.

\bibitem{mellia03}
Marco Mellia, Ion Stoica, and Hui Zhang.
\newblock {TCP}-aware packet marking in networks with diffserv support.
\newblock {\em Computer Networks}, 42(1):81--100, May 2003.

\bibitem{giacomazzi03}
G.~Verticale P.~Giacomazzi, L.~Musumeci.
\newblock {Transport of TCP/IP traffic over assured forwarding
  IP-differentiated services}.
\newblock {\em IEEE Network}, (5):18--28, September 2003.

\bibitem{padhye98modeling}
Jitedra Padhye, Victor Firoiu, Don Towsley, and Jim Kurose.
\newblock Modeling {TCP} throughput: {A} simple model and its empirical
  validation.
\newblock In {\em Proc. of {ACM SIGCOMM}}, pages 303--314, Vancouver, CA, 1998.

\bibitem{park04proportionnal}
Eun-Chan Park and Chong-Ho Choi.
\newblock {Proportional Bandwidth Allocation in DiffServ Networks}.
\newblock In {\em Proc. of {IEEE INFOCOM}}, Hong Kong, March 2004.

\bibitem{AQUILA}
Adaptive resource control for qos using an ip-based layered architecture.
\newblock http://www-st.inf.tu-dresden.de/aquila/.

\bibitem{GCAP}
Global communication architecture and protocols for new qos services over ipv6
  networks.
\newblock http://www.laas.fr/GCAP/.

\bibitem{GEANT}
GÉant : The pan-european gigabit research network.
\newblock http://www.dante.net/geant/.

\bibitem{TFTANT}
Tf-tant: Differentiated services testing.
\newblock http://www.dante.net/quantum/qtp/.

\bibitem{TEQUILA}
Traffic engineering for quality of service in the internet, at large scale.
\newblock http://www.ist-tequila.org/.

\bibitem{rfc3168}
K.~Ramakrishnan, S.~Floyd, and D.~Black.
\newblock The addition of explicit congestion notification ({ECN}) to ip.
\newblock Request For Comments 3168, {IETF}, September 2001.

\bibitem{sahu00achievable}
Sambit Sahu, Philippe Nain, Christophe Diot, Victor Firoiu, and Donald~F.
  Towsley.
\newblock On achievable service differentiation with token bucket marking for
  {TCP}.
\newblock In {\em Measurement and Modeling of Computer Systems}, pages 23--33,
  2000.

\bibitem{seddigh99bandwidth}
N.~Seddigh, B.~Nandy, and P.~Pieda.
\newblock {Bandwidth Assurance Issues for {TCP} Flows in a Differentiated
  Services Network}.
\newblock In {\em Proc. of {IEEE GLOBECOM}}, page~6, Rio De Janeiro, Brazil,
  December 1999.

\bibitem{yeom99realizing}
Ikjun Yeom and Narasimha Reddy.
\newblock Realizing throughput guarantees in a differentiated services network.
\newblock In {\em Proc. of IEEE International Conference on Multimedia
  Computing and Systems- ICMCS}, volume~2, pages 372--376, Florence, Italy,
  June 1999.

\bibitem{yeom01adaptive}
Ikjun Yeom and Narasimha Reddy.
\newblock Adaptive marking for aggregated flows.
\newblock In {\em Proc. of {IEEE GLOBECOM}}, San Antonio, Texas, USA, November
  2001.

\bibitem{yeom01modeling}
Ikjun Yeom and Narasimha Reddy.
\newblock Modeling {TCP} behavior in a differentiated services network.
\newblock {\em IEEE/ACM Transactions on Networking}, 9(1):31--46, 2001.

\end{thebibliography}

\end{document}